\input mn.tex
\input psfig.tex
\loadboldmathnames

\def\centreline{\centerline}
\def\etal{et~al.~}
\def\wtheta{w(\theta)}
\def\wqg{w_{qg}(\theta)}
\def\wgg{w_{gg}(\theta)}

\def\Aqg{A_{qg}}
\def\Agg{A_{gg}}
\def\Bqg{B_{qg}}
\def\Bgg{B_{gg}}
\def\simge{ \mathrel{\rlap{\raise 0.511ex \hbox{$>$}}{\lower 0.511ex \hbox{$\sim$}}}}
\def\simle{ \mathrel{\rlap{\raise 0.511ex \hbox{$<$}}{\lower 0.511ex \hbox{$\sim$}}}}
\def\point{{\cdot}}

\pageoffset{-2pc}{0pc}
\begintopmatter
\title{The environments of intermediate-redshift QSOs: $0\point3<z<0\point7$}
\author{R.~J.~Smith$^{1,2}$, B.~J.~Boyle$^3$ and S.~J.~Maddox$^{1,4}$}
\affiliation{$^1$ Institute of Astronomy, University of Cambridge, Madingley Road, Cambridge, CB3 0HA, UK}
\affiliation{$^2$ Research School of Astronomy and Astrophysics - Mount Stromlo Observatory, Institute of Advanced Studies, Australian National University, Private bag, Weston Creek P.O., ACT 2611, Australia}
\affiliation{$^3$ Anglo-Australian Observatory, PO Box 296, Epping, NSW 2121, Australia}
\affiliation{$^4$ School of Physics \& Astronomy, University of Nottingham, University Park, Nottingham NG7 2RD, UK}
\shortauthor{R.J.~Smith, B.J.~Boyle and S.J.~Maddox}
\shorttitle{Environments of intermediate-redshift QSOs}

\abstract{
An angular correlation of low significance ($2\sigma$)
is observed between $0\point 3<z <0\point5$ QSOs and $V\le 23$
galaxies.  Overall, the cross-correlation function between 
82~intermediate-redshift ($0\point 3<z<0\point 7$), X-ray selected QSOs and
$V\simle 24$ galaxies is investigated, but no signal is detected for the $z>0\point5$
QSOs. After converting to an excess of galaxies physically associated
with the QSO, this lack of strong correlation is shown to be
consistent with the clustering of normal galaxies at the
same moderate redshifts. Combined with previous observations, these results
imply that the environments of radio-quiet QSOs do
not undergo significant evolution with respect to the galaxy population
over a wide range of redshifts
($0<z<1\point5$). This is in marked contrast to the rapid increase in the richness
of the environments associated with radio-loud QSOs over the same
redshift range. 
}

\maketitle

\section{Introduction}
The study of QSO environments provides a direct measure
of the bias in their distribution with respect to galaxies. 
This not only provides insights into the triggering/fuelling mechanism
of QSOs, but assists in the relating the results from QSO
clustering studies (see e.g., Croom \& Shanks 1996, La Franca, Andreani \& 
Cristiani 1998) to large-scale structure studies carried out with
galaxy redshift surveys.

We have already investigated the galaxy cluster environments of
low-redshift QSOs (Smith, Boyle \& Maddox 1995, hereinafter Paper~I).
The results of this study demonstrated that radio-quiet QSOs exist in
average galaxy environments, consistent with these QSOs and galaxies
having the same clustering amplitude.
However, the observed objects were all at low redshift
($z\le0\point3$), and at low luminosity ($M_B>-24$).  In this paper we
extend this work to higher redshifts and higher luminosities,
probing the region of $M_B,z$ space where the forthcoming
large QSO surveys (2dF, Sloan) are targetted.
  
Previous surveys have been limited by relatively small samples of
radio-quiet objects, particularly in the redshift range $0\point3<z<0\point7$.
Yee and collaborators (Yee \& Green 1984, 1987, Ellingson, Yee \&
Green 1991) have previously studied the environments of predominantly
radio-loud QSOs at $z<0\point6$ and noted significant evolution in the
environments of QSOs towards rich clusters at higher redshift.  For
radio-quiet QSOs Ellingson et al.\ (1991) measure an amplitude of
$1\point1\pm0\point6$ times the mean, zero-redshift galaxy
autocorrelation amplitude and see no evidence for evolution over the
redshift range, $0\point25<z<0\point6$.  At higher redshifts
($1<z<1\point5$), most studies (Boyle \& Couch 1993, Croom \& Shanks 1998)
find no significant clustering around radio-quiet QSOs.

In comparison to earlier studies,
the advantages of the present work are in the increased number of
fields observed and the homogeneous, X-ray selection of QSOs. We are
also assisted by having larger and fainter galaxy redshift surveys
available with which to compare our results.

In Section~2 we describe the data and its reduction.
Section~3 summarises the techniques for measurement
of the angular clustering amplitude and the estimation from this
of a spatial clustering amplitude. Section~4 describes and discusses
the results and a summary is presented in Section~5.

Except where explicitly stated to the contrary, an Einstein--de Sitter
Universe has been assumed throughout. That is, $\Omega_0=1$,
$q_0=0\point5$ and $\Lambda_0=0$.  The Hubble constant is taken as
$100h\,$km\,s$^{-1}$\,Mpc$^{-1}$.

\section{Data reduction}
\subsection{Observations}
We used deep $V\/$-band CCD images to detect faint ($V<24$)
galaxies around QSOs in the redshift range $0\point3<z< 0\point7$.  QSOs were
drawn from the EMSS (Stocke \etal1991) and CRSS (Boyle \etal1997)
serendipity surveys which use {\sl Einstein\/} and {\sl ROSAT\/} data
respectively.  Details of the EMSS data and the significance of X-ray
selection are covered in Paper~I. Here we simply note that X-ray
selection ensures a fair sample of the general QSO population (Avni \&
Tananbaum 1986). It is largely insensitive to bias in radio luminosity
and spectral index (Zamorani \etal1981, Della Ceca \etal1994) and to
moderate amounts of dust extinction (Boyle \& Di Matteo 1995) that can
in principle seriously affect optical selection techniques.

$V$\/-band CCD images were obtained of 10-arcminute fields around 82~X-ray
selected AGNs in the redshift range, $0\point 3<z<0\point 7$ during
two observing runs (17--23/09/95 \& 15--21/04/96) at the $2\point5$-metre Isaac Newton 
Telescope (INT) on La Palma. The same $1024\times 1024$ thinned {\sl Tek3} CCD
camera was used on both occasions. The {\sl Tek3} chip has a peak
quantum efficiency of 75 per cent at 6500{\AA} (${>}$60 per cent
throughout 4000--7700\AA) and mounted at the prime focus of the INT, a
pixel scale of $0\point5896\,$arcsec$\,$pixel$^{-1}$. The
field-of-view corresponds to $2\point1\,h^{-1}$Mpc at $z=0\point5$.  The CCD was
read out in `Quick' mode; read noise $5\point 9$ electrons, gain
$1\point 44$ electrons$\,$ADU$^{-1}$. Exposures varied from
$15\,$minutes to $1\,$hour depending on the QSO's redshift and seeing
conditions at the time. In total 69 EMSS and 13 CRSS QSOs were
observed. The redshift and absolute magnitude distributions of the
sample are shown in Fig.~1.

\beginfigure{1}
  \centreline{\psfig{figure=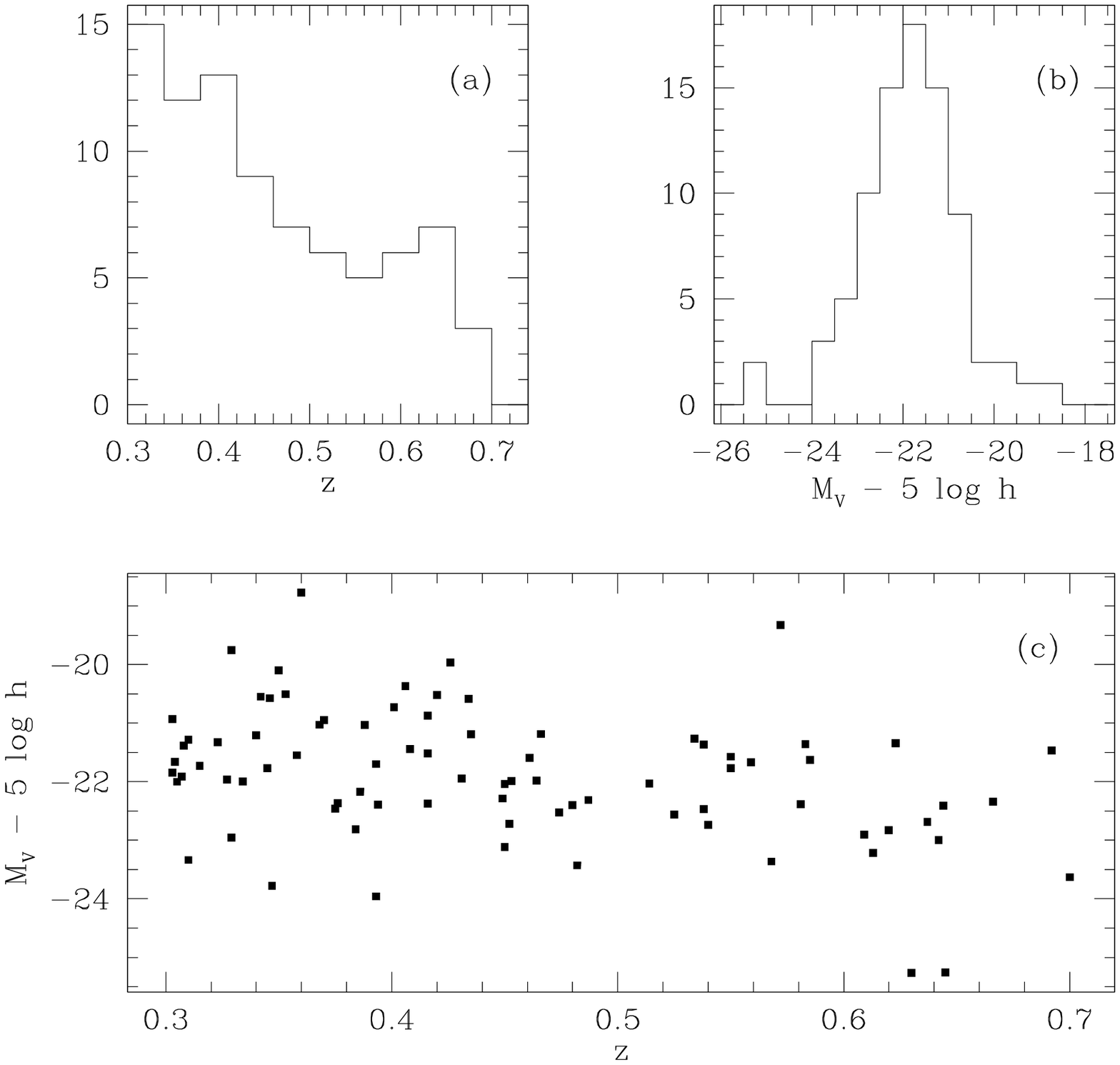,height=\hsize}}
  \caption{{\bf Figure  1.}
    (a) Redshift histogram of the 82 QSOs in the sample.
    (b) Absolute magnitude histogram. (c) Redshift and absolute magnitude are
    weakly correlated, a natural consequence of both evolution in the QSO luminosity
    function and this being a flux-limited sample.}
\endfigure

\subsection{Source detection}
Raw images were bias subtracted, flat-fielded and coadded
using the {\tt imred.ccdred} package in {\tt IRAF}. Source detection,
extraction, classification and photometry were carried out using
the {\tt SExtractor} software (Bertin \& Arnouts 1996). Each image was convolved with a Gaussian 
filter with full width at half maximum (fwhm) equal to the image's seeing,
a background map determined and potential sources selected as
connected groups of at least five pixels, each of which were 
individually greater than $1\sigma$ above the background. For more details of the 
detection and extraction process, see Bertin \& Arnouts (1996).

Most fields contained saturated stars, artificial satellite trails and
other defects. Such objects make detection of faint galaxies in their
immediate vicinity difficult, so these regions were excised from the
survey area in a semi-automated manner (see Smith 1998 for details).
As an illustration, the holes used for field MS01084$+$383, a very low
Galactic latitude ($b=-24^\circ$) field significantly affected by
bright stars, are shown in Fig.~2a.

\beginfigure*{2}
  \vfill
  \centreline{\line{
    \psfig{figure=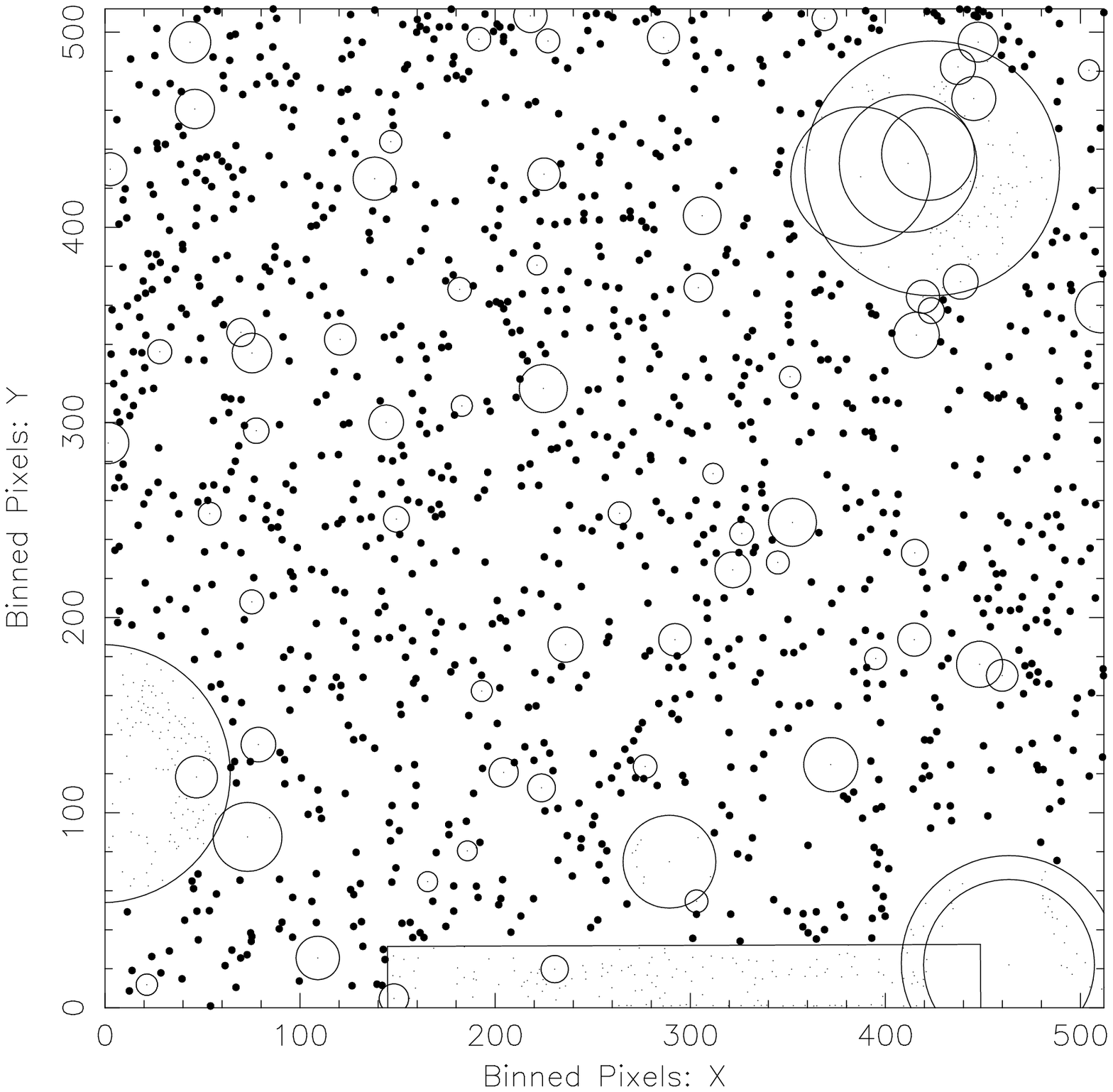,angle=-90,height=3.00in}\hfil
    \psfig{figure=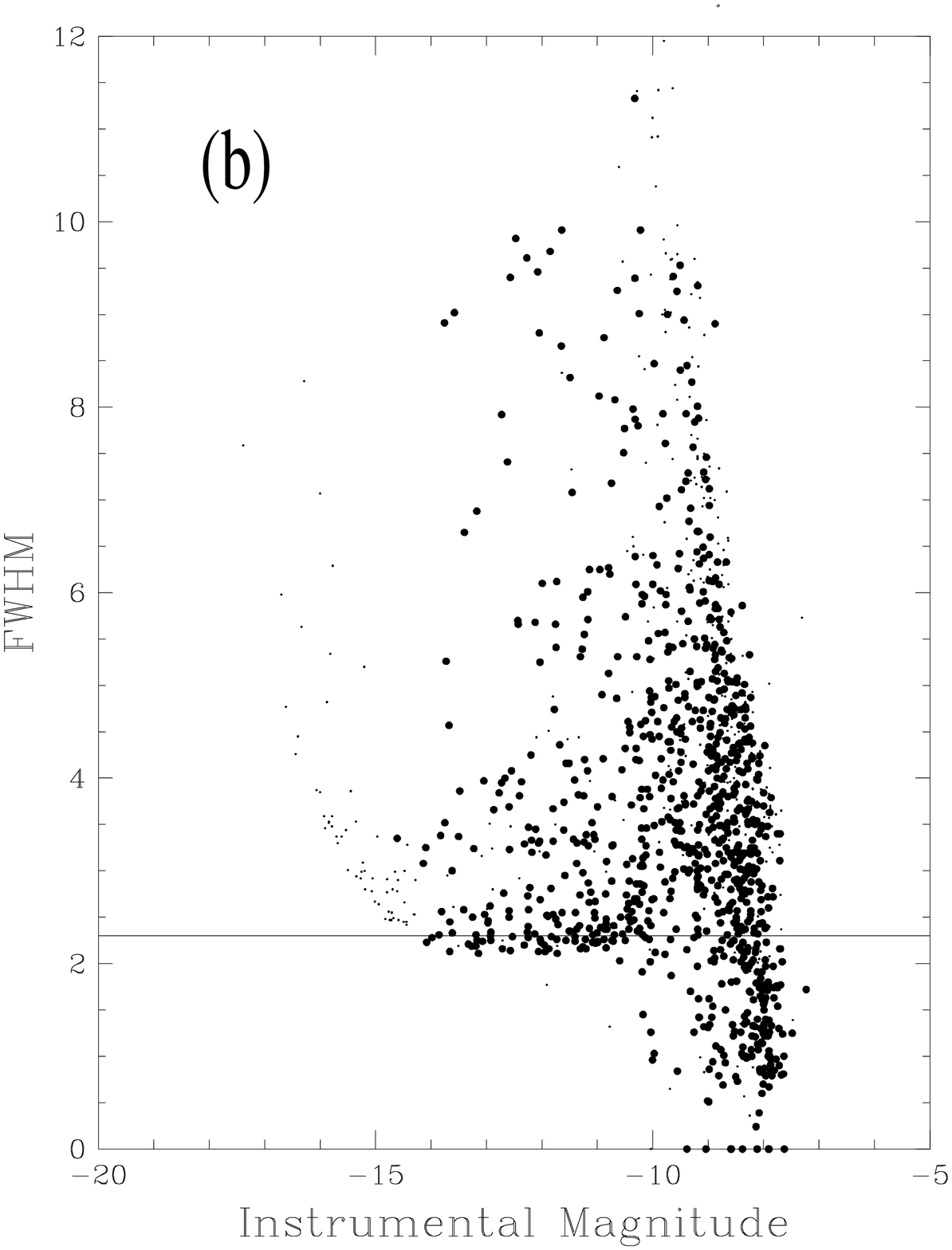,angle=0,height=3.00in,width=3.00in}\hfil
  }}\medskip
  \centreline{\line{
    \psfig{figure=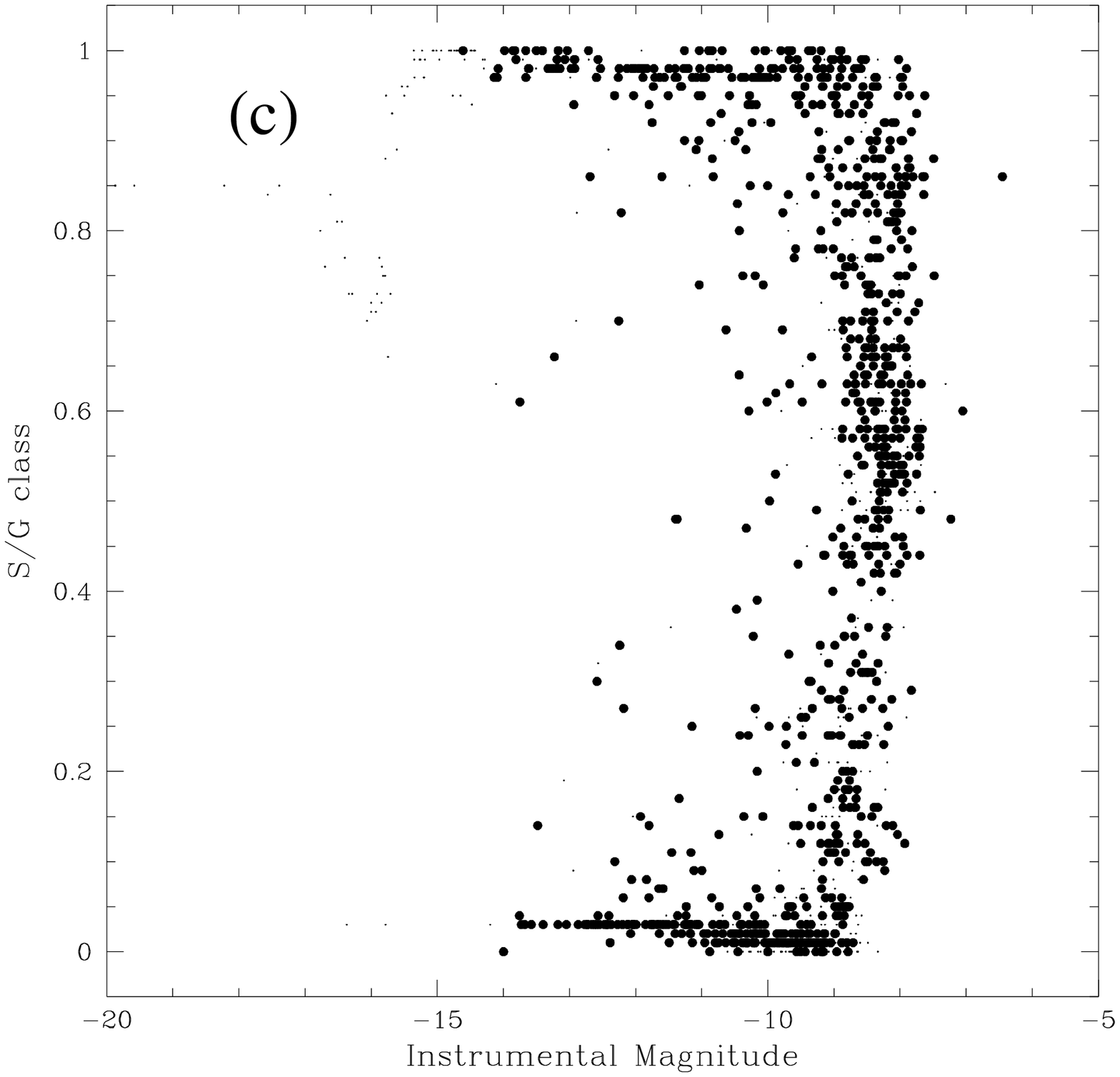,height=3.00in}\hfil
    \psfig{figure=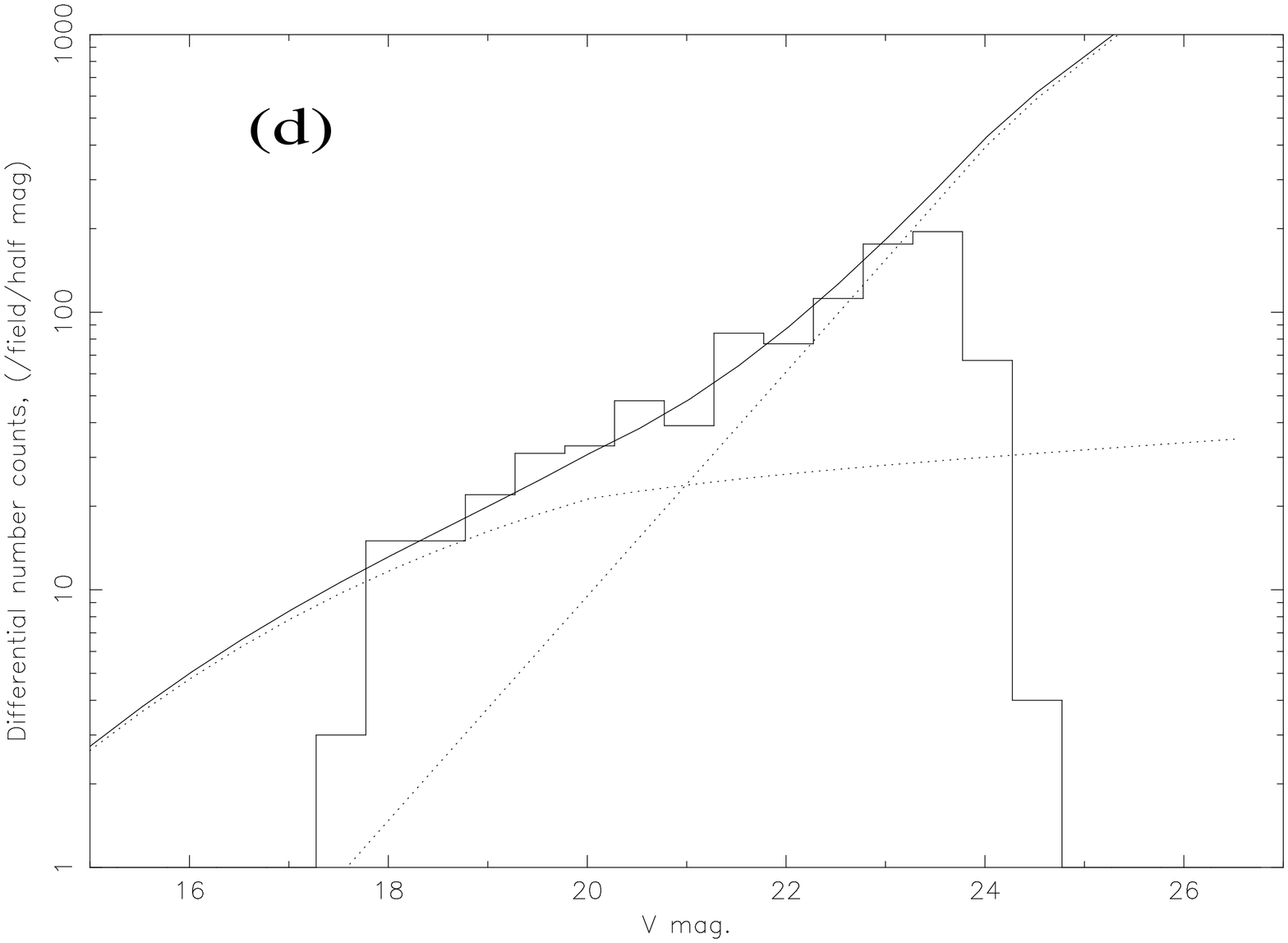,angle=-90,height=3.00in,width=3.00in}\hfil
  }}
  \caption{{\bf Figure 2.} 
  Four plots showing aspects of the reduction and photometric calibration
  process for one particular field, MS01084$+$383. 
  From the top left: 
  (a) All the sources detected in the field.
  The $1024\times1024\,$pixel {\sl Tek} CCD has been re-binned to $512\times512\,$pixels 
  for display purposes only. Large circles and squares show regions of the frame excised 
  for any reason, usually proximity to a saturated star.
  (b) Determination of the seeing parameter.  The horizontal line shows
  the selected value for the stellar fwhm in pixels. 
  (c) A scatter plot of the star--galaxy classification flag against instrumental 
  magnitude. A value of unity signifies a star and zero a galaxy. Values between
  these extremes lie on a non-linear probability function of being stellar. 
  In all three plots, large points show sources included in the final catalogue
  while small dots are objects which fall in regions of the
  frame excised for any reason.
  Note that saturated stars deviate significantly from a point-source profile (panel b)
  and are successfully rejected (appear as small dots) rather than mis-classified
  as galaxies (panel c).
  The final panel (d) shows determination of the photometric zero-point. The solid 
  curve is the expected number counts and the histogram the observed counts. Calibration
  consists of sliding the histogram horizontally until the best fitting position is found.
  The dotted lines are the expected counts for stars and galaxies plotted separately.  }
\endfigure

To optimise source detection and classification, it is necessary to
know the seeing a priori. This information is used to choose a
convolution filter for each frame and as the single input parameter
into the star--galaxy separation algorithm (see below and Bertin \&
Arnouts 1996).  The {\tt SExtractor} software was therefore run on
each frame twice. After the first run, the output catalogue contains
the fwhm of every object in the frame allowing an accurate assessment
of the seeing, which was taken as the modal fwhm value for unsaturated
images. See Fig~2b.  These seeing estimates were inspected
individually. In a few cases with poor seeing the modal fwhm was not a
robust estimate because of a broad scatter in the stellar locus.  For
these the seeing was estimated by inspection of the data frame and
measuring individual clearly stellar objects. In following this
procedure it was consistently shown that more objects were detected in
the second pass and the star--galaxy separation was more
reliable. It was also confirmed that the second pass did not
alter the assessment of the seeing.

\subsection{Photometric calibration}
Weather conditions during the observing runs did not merit calibration
by photometric standard stars, so the CCD images were calibrated 
by comparison with previously published, photometrically calibrated deep 
number--magnitude counts. We used the deep $V\/$-band galaxy counts of 
Smail \etal(1995) and we took the stellar counts from the Galaxy model 
of Bahcall \& Soneira (1980).

By fitting the instrumental magnitude number counts from the CCD data
to the star-plus-galaxy counts expected on the
basis of the chosen models, the only free parameter
was the photometric zero-point. The best fit was defined as that giving
the minimum absolute deviation between the published and instrumental
counts between $V=20$ and $0\point 5\,$magnitude brighter than the
peak in the instrumental counts.  The observed number counts were
generally found to maintain an excellent fit to the model counts at
$V<20$ until the CCD saturated at $V\approx18$.  The full catalogue
was used in order to prevent errors in the star--galaxy separation
from propagating into the calibration.  One such calibration plot is
shown in Fig.~2d.  The observed number counts are shown as a
histogram. The expected counts are shown as the solid curve, with the
stellar and galaxian counts plotted individually as dotted
curves. Note that this is a low Galactic latitude ($b=-24^\circ$)
field and the inclusion of the stellar counts
becomes very important. For high Galactic latitude fields, the Smail
et~al.  galaxy counts alone provide an adequate fit to the shape of
the observed counts.

This method of calibration is subject to intrinsic variations in the
source number density, which are small at faint limits, and to the
Poisson error on the counts.  Assuming the $b_J\le24$ galaxy
auto-correlation of Roche \etal(1993) and integrating over the CCD
field of view, intrinsic galaxy clustering introduces an error of less
than $0\point01\,$mag.  The star counts model is only claimed to have a
15 per cent accuracy, but since stars typically constitute much less than five
per cent of the sources, they also only introduce an error of order $0\point01\,$mag.
In the very low Galactic latitude fields, it may be as much as $0\point03\,$mag.
With at least five hundred sources used in the calibration for each field, 
the Poisson error introduces an uncertainty of less than $0\point05\,$mag.

\subsection{Star--galaxy separation}
The fields investigated cover a wide range of Galactic
latitudes. Whilst the majority have latitudes $|b|\ge40^\circ$, there
are a significant number nearer the Galactic plane and eight fields
were observed at as low as $|b|<30^\circ$. Stellar contamination of
the galaxy catalogue will reduce the amplitude of any observed
correlation (see Section~3.3).  At low Galactic latitudes in particular, removal of the
stars is therefore important. This may be done by excising the stars
from the catalogue before estimating the correlation function, or by
estimating the amount of contamination and re-scaling the correlation
function to account for it.  Both approaches were followed producing
results entirely consistent within the statistical errors. The results
quoted herein are based on the procedure that used the star--galaxy classification.

The classification as stars or galaxies was achieved using the
algorithm in the {\tt SExtractor} package, which
produces a classification flag with a value lying between 0 (extended) and 1 (compact). 
A value of 0$\point$8 was chosen as the boundary between stellar and galaxian 
images by inspection of plots such as Fig.~2c. This value 
includes the full width of the locus of points
at $\sim 1$ while excluding the cluster which frequently 
appears at $\sim 0\point5$. Such a cluster is seen only weakly in the example shown here,
but is much more pronounced for some fields. The algorithm's defaulting
to values $\sim 0\point5$ is effectively a {\sl don't-know\/} classification, but
our cut of 0$\point$8 means such objects are called galaxies.

Checks of the validity of the star--galaxy separation process, including
the generation of artificial data using the {\tt artdata} package in {\tt IRAF}
demonstrated that the star--galaxy classification procedure gave
reliable results (see Smith 1998).

\subsection{The minimum observable angular scale}

Artificial data generated by the {\tt artdata} package was also used
to establish the effect of the bright QSO on detection of nearby faint
galaxies. In order to do this $10\,000$ small ($100\times100$ pixels)
simulations were carried out using input parameters selected randomly
from distributions defined by the complete CCD data set. The seeing,
zero point, QSO magnitude and background level were all varied in this
way (see Smith 1998 for details). 
Each image was furnished with sources down to one magnitude fainter
than the CCD completeness and then extracted and identified as
usual. There was a marked failure to find galaxies within a few
pixels of the modelled QSO. The ratio between the number of simulated 
and detected sources was computed as a function of 
radius from the QSO. At distances greater than sixteen pixels, the bright saturated
object no longer had a detrimental effect and the source detection success rate 
was constant, but it decreased rapidly nearer the QSO.  At less than five pixels from the
QSO, the source detection rate was virtually zero.  No counts were therefore performed closer than
5 pixels to the QSO. A best fitting, smooth polynomial was determined to give a 
correction factor for counts at separations between 5 and 16 pixels.  
Though locally very significant, the 5 to 16 pixel annulus is very small,
so its overall impact on observed counts is also small and makes only a
quantitative, not qualitative difference to the results. 
For example, counts out to one arcminute required multiplication by $1\point0053$
to compensate for the erroneously low counts in the central few arcseconds.

\section{Analysis}
\subsection{The correlation function}
The QSO--galaxy angular correlation function, $\wqg$, measures the
statistical excess of galaxies observed near the QSO over that
expected for a random distribution of the same number of galaxies. In
its standard form it may be defined
$$n(\theta)\delta\Omega = N_g[1+\wqg]\delta\Omega, \eqno\stepeq$$
where $n(\theta)\delta\Omega$ is the number of galaxies observed in
solid angle $\delta\Omega$ at angular separation $\theta$ from the
QSO. $N_g$ is the background surface density of galaxies.  The galaxy
auto-correlation function may be defined similarly as the joint
probability of finding a galaxy in both solid angle elements
$\delta\Omega_1$ and $\delta\Omega_2$ separated by angle $\theta$.
$$\delta P(\theta)=N_g[1+\wgg]\delta\Omega_1\delta\Omega_2
\eqno\stepeq$$ The galaxy angular correlation function has been
extensively studied (e.g., Groth \& Peebles 1977, Maddox \etal1990)
and found to be well fit at scales less than $5^\circ$ by a power-law
of the form
$$w_{gg}(\theta)=A_{gg}\theta^{1-\gamma}, \eqno\stepeq$$ where
$A_{gg}$ is the angular covariance amplitude. The power-law slope is
measured to be in the range $1\point7<\gamma<1\point8$.  We here
assume the QSO--galaxy correlation function to be of the same form,
$$\wqg=\Aqg\theta^{1-\gamma} \eqno\stepeq$$
We define $\Aqg$ to be the amplitude at one degree except where explicitly
stated to the contrary.

Measurement of the angular covariance amplitude alone does not tell us
about the physical environment of the QSO. We are integrating along
the line of sight to the QSO, including information about the
foreground and background galaxy populations.  The more physically
relevant spatial covariance amplitude, $\Bqg$, gives the strength of
spatial clustering of galaxies actually associated with the QSO. The
approach adopted here is to compare this with the galaxies' own
clustering amplitude, $\Bgg$, to determine the relative environmental
properties of QSOs and galaxies.

By analogy to equation~1, the spatial cross-correlation function,
$\xi_{qg}(r)$, is defined as
$$n(r)\delta V = \rho_g[1+\xi_{qg}(r)]\delta V,\eqno\stepeq$$ where
$n(r)\delta V$ are the counts in volume $\delta V$ at distance $r$
from the QSO and $\rho_g$ is the average volume density of
galaxies. The galaxy auto-correlation function is similarly defined as
$$\delta P(\theta)=\rho_g^2[1+\xi_{gg}(r)]\delta V_1\delta V_2 \eqno\stepeq$$

Limber's Equation (Limber 1953) relates the amplitudes of the angular
and spatial auto-correlations and shows that equation~3 is a natural
consequence of spatial clustering of the form
$$\xi_{gg}(r)=B_{gg}r^{-\gamma}=\left({r\over
r_{0,gg}}\right)^{-\gamma}. \eqno\stepeq$$ Measurements are quoted in
the literature either as the spatial covariance amplitude $\Bgg$, or
as a correlation length $r_{0,gg}$. QSO--galaxy cross-correlations
have not been measured to sufficient precision to allow a free fit,
but are generally assumed to follow the same simple power-law
form. (e.g., Yee \& Green 1987).
$$\xi_{qg}(r)=B_{qg}r^{-\gamma}=\left({r\over r_{0,qg}}\right)^{-\gamma}. \eqno\stepeq$$
 
Longair \& Seldner (1979), hereinafter LS79, showed how to convert an
observed $\Aqg$ to an implied $\Bqg$.  A similar approach is adopted
here with minor modifications.  The conversion is performed in a
statistical sense, since the galaxy redshifts 
are not available. We therefore make several well-justified
assumptions relating to the redshift distribution of galaxies and the
form and evolution of the galaxy clustering.

First, all clusters are taken to be spherically symmetric. This may
not be so for individual cases, but is reasonable when averaged over a
large sample. Secondly, we need to allow for the redshift distribution
of galaxies, though not the actual redshift of any particular object.
Previous applications of this technique have predominantly derived an
$N(z)$ from a fixed luminosity function (LS79, Yates, Miller \& Peacock 1989, 
Boyle \& Couch 1993).  Though it is possible to improve on this by using an evolving
luminosity function (Ellis \etal1996, Maddox \etal1990), we have here
adopted the parametric fit to observational redshift-survey data by
Efstathiou (1995).  (See also Baugh \& Efstathiou 1993, Maddox, Efstathiou \& Sutherland 1996.)  
Deep redshift surveys (e.g., Glazebrook \etal1995) and studies of weak
lensing around rich galaxy clusters (Smail, Ellis \& Fitchett 1994)
mean the $N(z)$ relation for $B<24$ galaxies is now well determined,
and this offers a more directly observational route to obtaining the
required data.
The parametric form for the galaxy number-redshift relation derived
by Efstathiou (1995) is given by,
$$
\eqalignno{
  {{\rm d}N\over {\rm d}z}&\propto z^2\exp\left\{ -\left[ {z\over z_c(b_J)}\right]^{3/2}\right\}&\startsubeq\cr
  z_c(b_J)&=\cases{0\point0113(b_J-17)^{1\point5}+0\point0325& $17 \le b_J \le 22$\cr
                    0\point0010(b_J-17)^{3}+0\point0325& $b_J > 22$\cr}&\stepsubeq\cr
}
$$
The relation was normalised assuming a mean galaxy $b_J-V = 0\point5,$
from the same Smail \etal(1995) $V\/$-band counts as used for the
photometric calibration in Section~2.3. The mean galaxy $B-V$ colour
was derived from the data of Fukugita, Shimasaku \& Ichikawa (1995)
and Driver \etal(1994), using the morphology mix shown in Table~1 (see
Prestage \& Peacock 1988), and converted to $b_J-V$ via the colour
equations of Blair \& Gilmore (1982).  The redshift
distributions used are shown in Fig.~3.

\begintable{1}
\caption {{\bf Table 1.}
  Assumed galaxy morphology mix used in derivation of
  mean galaxy colours and K-corrections.}
\centreline{
\vbox{
\tabskip=1em plus 2em minus .5em
\halign{\hfil#\hfil&\hfil#\hfil\cr
\noalign{\hrule}
\noalign{\smallskip}
\noalign{\hrule}
\noalign{\medskip}
Galaxy class&Fractional abundance\cr
\noalign{\smallskip}
\noalign{\hrule}
\noalign{\medskip}
E / S0 & $0\point35$\cr
Sab & $0\point 20$\cr
Sbc & $0\point 20$\cr
Scd & $0\point 15$\cr
Im & $0\point 10$\cr
\noalign{\medskip}
\noalign{\hrule}
} }
}
\endtable

\beginfigure{3}
  \centreline{\psfig{figure=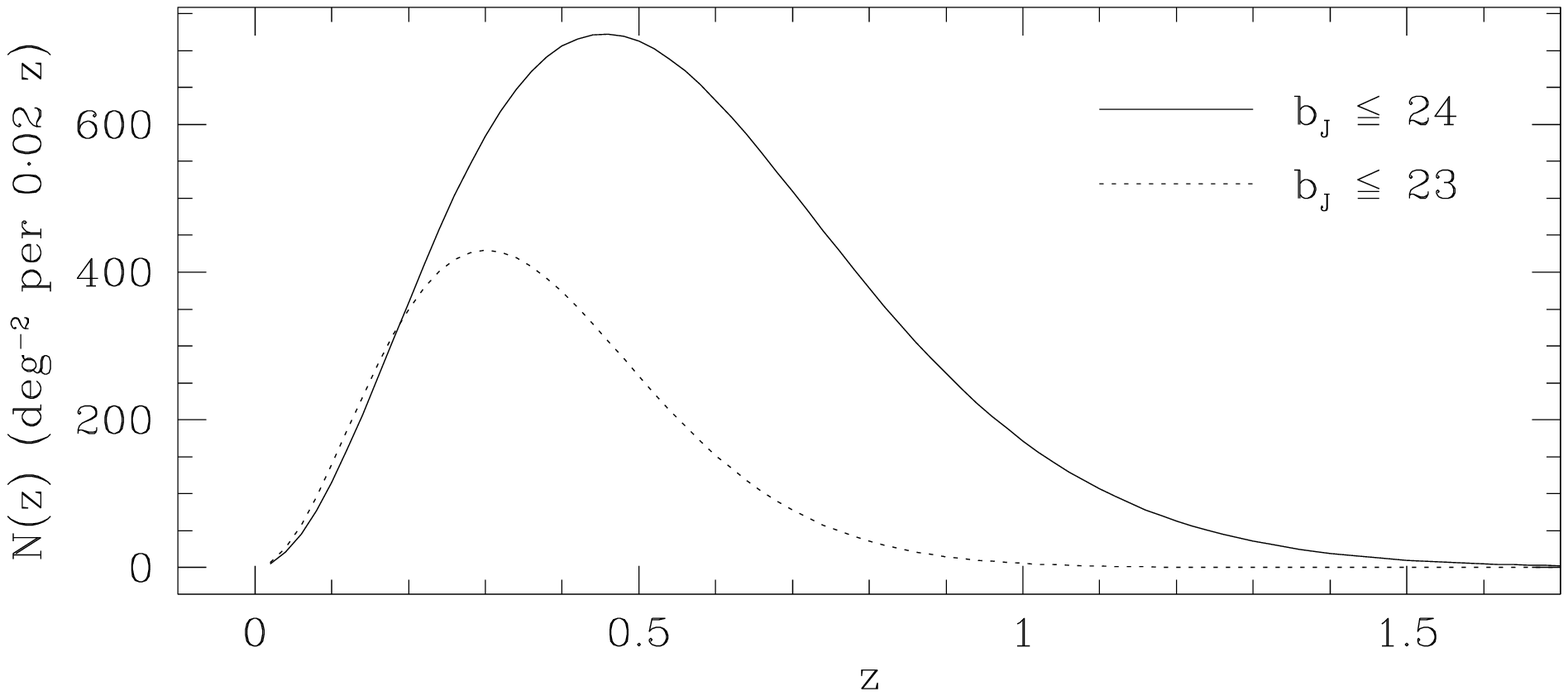,width=\hsize}}
  \caption{{\bf Figure 3.}
     The galaxy redshift distributions adopted for the de-projection
     of the observed angular cross-correlation. They were derived from
     the functional form given by Efstathiou (1995). These may be compared with
     the QSO redshift distribution shown in Fig.~1.}
\endfigure

Finally LS79 assumed that the clustering is stable in proper
coordinates. This is the case for virialized clusters where the
structures behave as particles within the expanding Universe. On
larger scales or for younger, non-virialized clusters, there is an
intrinsic expansion or contraction within the cluster superimposed on
the Hubble expansion.  For a cluster simply expanding with the Hubble
flow, the clustering signal is constant in comoving coordinates.  
This is expected in the case of sources that have a constant high
bias, where the distribution of sources is determined by the initial
conditions, largely independent of dynamical evolution.

The rate of clustering evolution is here included as a free parameter,
which makes the following changes to the derivation in LS79.
Following Phillipps \etal(1978), we assume that only the amplitude
varies with redshift, the form remaining constant.  Equation~8
becomes,
$$\xi_{qg}(r,z)=B_{0,qg} f(z) r^{-\gamma}.\eqno\stepeq$$ 
Let the typical cluster scale length, $R$, evolve as $R(z) \propto
(1+z)^{-\beta}.$ From the definition of the correlation function
(equation~1), the excess number of galaxies, $N_e$, in a shell around
the QSO is
$$N_e=\rho_g(z) \delta V(z) \xi_{qg}(R(z),z).\eqno\stepeq$$ 
This shell is then scaled with redshift according to the proposed
evolution model. For observations to progressively higher redshifts,
the volume sampled changes, making $\rho_g(z)\propto(1+z)^3$ and
$\delta V \propto R^3 \propto (1+z)^{-3\beta}.$ Thus,
$$N_e \propto (1+z)^3 (1+z)^{-3\beta} (f(z)
(1+z)^{\beta\gamma}).\eqno\stepeq$$ 
The first two terms arise from the change in volume sampled.  The
final term, which may be seen by substituting $R(z)$ for $r$ in
equation~10, expresses the change in physical scale sampled for the
same comoving scale as a function of redshift.  Assuming the number of
galaxies physically in the cluster is a constant, $N_e$ is a constant,
and re-arranging, gives $f(z)
\propto (1+z)^{\beta(3-\gamma)-3}.$ Equation~10 is therefore
re-written as
$$\xi_{qg}(r,z)=B_{0,qg} (1+z)^\nu r^{-\gamma}; \qquad \nu = \beta(3-\gamma)-3. \eqno\stepeq$$

We substitute this form of the clustering evolution in place of that
used by LS79, thus slightly modifying their equation expressing
$\wtheta$ in terms of spatial parameters.
The final expression relating angular and spatial amplitudes is then
$${\Aqg\over B_{0,qg}} = {I_\gamma \over N_g} D^{3-\gamma} (1+z)^{\nu+\gamma} \varphi(m_0,z). \eqno\stepeq$$
The only difference between this and the LS79 equation is the power to which the 
redshift dependent term is raised. $I_\gamma$ is a constant dependent only on the choice
of slope for the power law in equation~10 and $D$ is the coordinate distance
to the QSO.
$$D = {{cz}\over{H_0}(1+z)}\left[{ {1+z+\sqrt{1+2q_0z}}\over{1+q_0z+\sqrt{1+2q_0z}} }\right]\eqno\stepeq$$
$\varphi(m_0,z)$ is an integral luminosity function, i.e., the number
of galaxies per comoving cubic mega-parsec with apparent magnitude brighter than
the flux limit of the sample, $m_0$, at redshift, $z$. As described above,
$\varphi$ may be expressed as ${\rm d}N/{\rm d}V$, the ratio of an observed ${\rm d}N/{\rm d}z$
and ${\rm d}V/{\rm d}z$, which is precisely specified by the chosen cosmology.

The value of $\beta$ then specifies the type of clustering evolution which is present.
If $\beta=1$, $\nu=-\gamma$ and the redshift dependence disappears giving comoving evolution. 
If $\beta=0$, then $\nu=-3$ and the redshift term becomes
$(1+z)^{-3+\gamma}$, which is the case for stable clustering as in LS79.
This evolution has also been parameterised in the literature in terms 
of the equivalent term $\epsilon$, where $\epsilon=\beta(\gamma-3)$ and
$\nu=-(3+\epsilon)$ (e.g., Efstathiou \etal1991).

\subsection{Practical estimators for $\bmath{\wqg}$ and $\bmath{\Aqg}$}
For a single field, the QSO--galaxy cross-correlation may be estimated by
$$\wqg = {QG\over QR} - 1,\eqno\stepeq$$
where $QG$ is the number of galaxies found in the range of separations
$(\theta-\Delta\theta/2,\theta +\Delta\theta/2]$ from the QSO. $QR$ is
the number of randomly scattered  points in the same annulus. 
Random points were generated within the survey boundaries taking
full account of any holes or complicated survey geometry.
In order to minimise random errors, 
many times as many random `galaxies' were generated as there
are real galaxies in the field, and the counts scaled appropriately.

The clustering signal from any single QSO is 
very weak and has large associated uncertainties due to both
Poisson errors and the intrinsic clustering of faint galaxies
unrelated to the QSO.  For most of the results
presented here, the counts were coadded for all the fields
within a given subsample (e.g., redshift slice), amounting
to the pair-weighted 
average of $j$ fields calculated individually, i.e.,
$$\wqg = {\sum^j_{i=1} QG_i \over \sum^j_{i=1} QR_i} - 1. \eqno\stepeq$$

The angular covariance amplitude was determined
in two ways. If the full correlation function is plotted for a
range of angular scales (e.g., Fig.~4), the best fitting value was found
for the assumed form (equation~4) of the corelation function.
Alternatively, if counts are taken in a single broad annulus around the 
QSO, the assumed form may be integrated to give a scaling factor
from the mean correlation in the annulus (the observed quantity)
to the amplitude at any chosen scale.  This is the approach adopted where
$\Aqg$ must be determined from a single QSO field, e.g., Fig.~7.
If $QG$ and $QR$ are the galaxy and random counts in the range 
$(\theta_{\rm min},\theta_{\vphantom{i}\rm max}]$, then integrating
equation~1 gives the excess counts above the background 
$$ 
\eqalignno{
QG-QR &= \int^{\theta_{\rm max}}_{\theta_{\rm min}}2\pi\theta N_g\wqg{\rm d}\theta&\startsubeq\cr
      &= {2\pi\theta N_gA_{qg} \over 3-\gamma}\left[{\theta^{3-\gamma}}\right]^{\theta_{\rm max}}_{\theta_{\rm min}},&\stepsubeq\cr }
$$
using our standard assumed functional form for $\wqg$.
The background count, $QR$, is $\pi N_g(\theta_{\vphantom{i}\rm max}^2 - \theta_{\rm min}^2)$, 
which substituted into the above equation gives,
$$A_{qg}={3-\gamma\over 2}\left({QG\over QR}-1\right)
  \left({\theta_{\vphantom{i}\rm max}^2 - \theta_{\rm min}^2 
  \over \theta_{\vphantom{i}\rm max}^{3-\gamma} - \theta_{\rm min}^{3-\gamma}}\right).\eqno\stepeq$$
If the angles are normalised such that $\theta_{\rm max}=1$,
this formula gives the amplitude at the outer rim of the annulus. If
they are expressed in degrees, it would be the amplitude at one degree and so forth.

\subsection{Effects of poor star--galaxy separation}
The correlation function thus far defined is
the {\sl true\/} correlation function, $w_t(\theta)$. In
practice the galaxy catalogue is likely to be contaminated by
stars or incomplete through the mis-classification of galaxies.
Since the background count is taken  from the total number of galaxies
in the field, both $QG$ and $QR$ are biased, which in turn biases the
{\sl observed\/} correlation function, $w_o(\theta)$.

In the case of stellar contamination leading to an over-estimate of the counts, 
it can straightforwardly be shown (e.g., Smith 1998) that: 
$$w_t(\theta)= w_o(\theta)\left(1+{{\cal N}_S\over {\cal N}_G}\right).\eqno\stepeq$$ 
${\cal N}_G$ and ${\cal N}_S$ are the total numbers of galaxies and stars 
respectively found in the field.
Since this is a multiplicative factor, the
power-law shape remains unchanged.  Neglecting to perform any
star--galaxy separation, can therefore be corrected by rescaling the
observed amplitude to its true value so long as the fractional
contamination by stars is known.

The alternative case is accidental exclusion of galaxies from the
catalogue.  Assuming any such failure is random, both $QG$ and $QR$
counts have the same fractional error and the observed correlation
function remains unchanged.  Fewer counts do of course imply a lower
statistical significance on any measurement. We note that the random
errors assumption may not be entirely precise.  Failures of the
star--galaxy separation are unlikely to be entirely uncorrelated,
perhaps having greater effect on galaxies in particular environments,
magnitude ranges or regions of the image. We have been unable to detect
any such systematic errors in the current data.

\subsection{The Integral constraint, $\bmath{I_B}$}
The true background number density of galaxies is not known precisely,
so was estimated from the data. The background density was initially
taken as the mean density within the particular field in question.  If
we then show that there is indeed a
statistical excess of galaxies in the field, the estimated background
value must have been biased high.  The error may be retrospectively
calculated from the observed correlation and the process iterated to
self-consistent values. The fractional error in the background, $I_B$,
or integral constraint, causes the true background to be
over-estimated by a factor $1+I_B$.  $I_B$ is determined by
integrating the correlation function over the full area of the field
used.
$$ I_B = {\vphantom{\displaystyle\int}{\rm Integrated\ excess\
         counts}\over\vphantom{\displaystyle\int}{\rm True\
         background}} = {{\displaystyle\int^{\theta_{\rm
         max}}_{\theta_{\rm min}}2\pi\theta\wtheta{\rm d}\theta}\over
         {\displaystyle\int^{\theta_{\rm max}}_{\theta_{\rm
         min}}2\pi\theta{\rm d}\theta}}\eqno\stepeq$$ 

assuming counts were taken from a circular region.  For more
complicated survey geometries, including holes in coverage, the
quantity $I_B$ is more easily obtained by numerical integration over
each field individually.

\subsection{Errors on $\bmath{w(\theta)}$}
Estimation of the errors on a correlation function and quantities
determined therefrom is a subject of active debate (e.g., Landy \&
Szalay 1993, Mo, Jing \& B\"orner 1992 and references therein).  The
measured amplitude of the auto-correlation function at any given scale is
dependent on the distribution of the galaxies over a wide range of
separations. Any individual galaxy will exist as one member of 
many different pairs, so the pair counts in any one counting
annulus are not independent of counts in the neighbouring annuli. As a
result, despite essentially being a counting statistic, the error on the pair counts is
not in general Poissonian.  The important distinction which must be
drawn here is between an auto-correlation or full cross-correlation of two catalogues and
the simplified geometry we are currently considering. This work simply
azimuthally averages the number density of galaxies as a function of
the radial distance from the QSO position.  As a result, the galaxies
in all pairs are independent so long as the fields do not overlap and
the galaxy distribution is fully defined by the QSO--galaxy
cross-correlation. In reality galaxies are clustered, so an assumption
that galaxies' positions are independent is not true. It is however a
very good approximation since any galaxy--galaxy clustering is washed
out around the counting annulus. The error on the galaxy counts should
then be very close to the $\sqrt N$ Poisson error.  This was confirmed
by a full bootstrap analysis (Barrow, Bhavsar \& Sonoda 1984) of the
errors (see Smith 1998) and so the Poisson error was adopted for this
analysis.

\section{Results}
\subsection{Full data set, $\bmath{0\point3 < \lowercase{z} < 0\point7}$}
The cross-correlation, $\wqg$, for all the QSOs with $V\le23$ galaxies
is shown in Fig.~4. There is a positive correlation, though 
detected at a moderately weak level.
At sub-arcminute separations there is a $2\sigma$ excess of
galaxies over that expected for a random distribution and for the smallest
counting annulus, approximately 22\,arcsec, it is greater than $3\sigma$.

The $V\le23$ cut was chosen to correspond to 
about one magnitude fainter than the characteristic absolute magnitude, 
$M_V^*-5\log h$, at $z\sim0\point 4$. It equals $M_V^*-5\log h$ at $z\sim0\point55$.
This ensures that the peak of the galaxy $N(z)$ relation lies 
in the same redshift range as the QSOs.  $M_V^*$, as a function of redshift, was taken
from the $b_J<24$ derived luminosity function of Ellis \etal(1996).
Galaxy K-corrections (Oke \& Sandage 1968) were taken from Pence (1976).
Values for different galaxy types were averaged according
to the Table~1 morphology mix and a linear fit made
over the range $0\point 3<z<0\point 7$. 

\beginfigure{4}
  \centreline{\psfig{figure=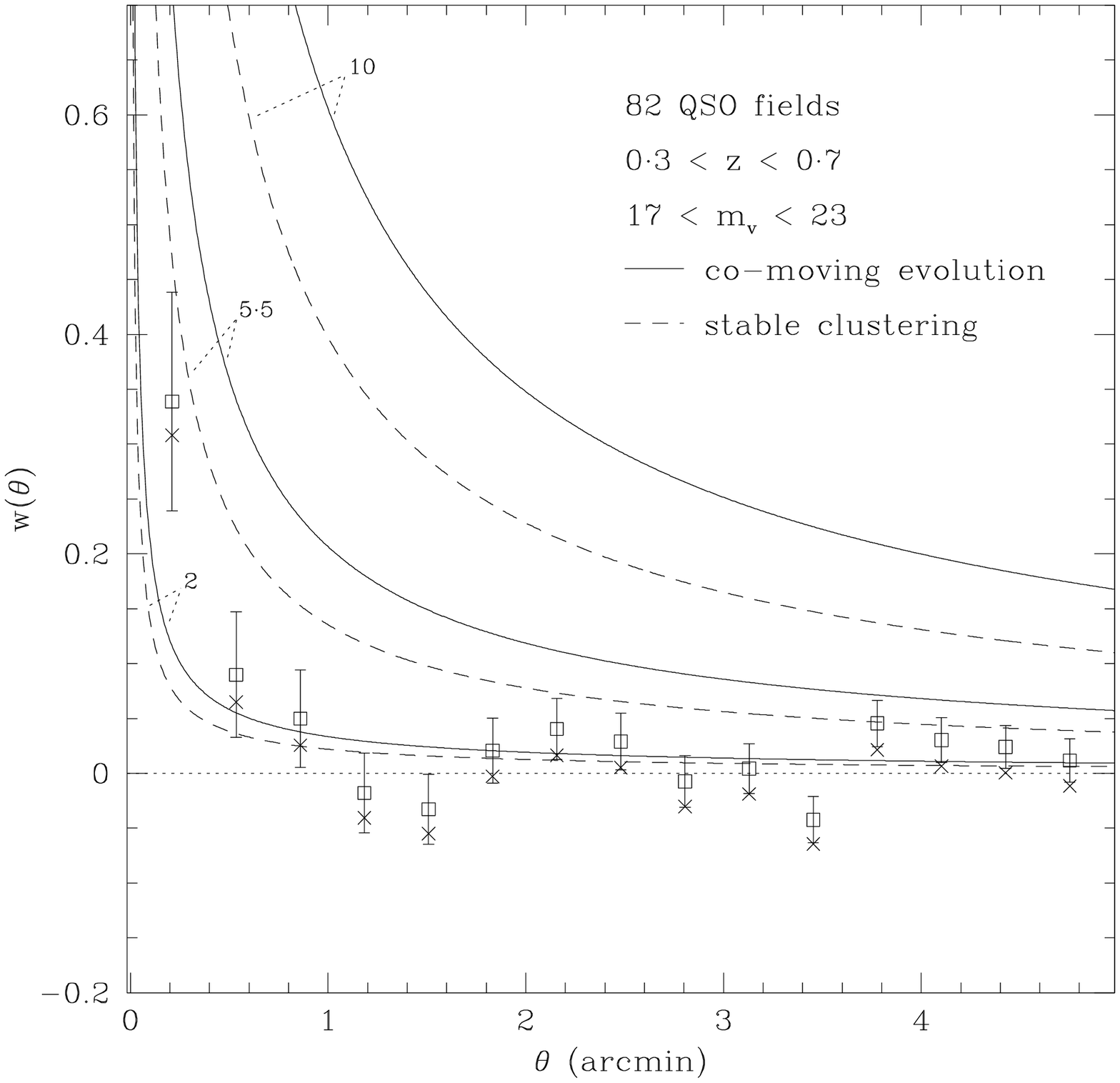,width=\hsize}}
  \caption{{\bf Figure 4.}
     The cross-correlation function for all 82
     $0\point3 < z < 0\point7$ QSOs with $V\le23$ galaxies.
     Crosses are the observed values. Open squares are modified by the integral 
     constraint ($I_B=11\point1 A_{qg}$). The six curves show the expected signal for
     correlation lengths of $r_0$ = 10, $5\point5$ and 2\,$h^{-1}$Mpc respectively
     from top to bottom. The solid  and dashed lines 
     represent $r_0$ as a constant in comoving and proper coordinates.}
\endfigure

The predicted correlation functions for six different clustering
amplitudes and evolution models have also been plotted in Fig.~4. Dashed
lines are used for models stable in proper coordinates ($\beta=0$, equation 13) and 
solid lines used where $r_0$ is assumed to be stable in comoving coordinates ($\beta=1$). 
To derive these curves, the correlation expected around each individual QSO was
calculated (equation~14) assuming a power-law slope of
$\gamma=1\point8$ and the mean of the 82 separate amplitudes (one for each
QSO field) is plotted. In this way, projection
effects due to the redshift distribution of the QSOs are accounted for. 
Whereas equation~14 is integrated over the generalised, mean galaxy $N(z)$,
this procedure effectively integrates over the $N(z)$ of the particular 
QSOs in the sample. The correlation lengths shown in Fig.~4 are
10, $5\point5$ and $2\,h^{-1}$Mpc, chosen to represent the following cases:
\beginlist
\item $r_0=2\,h^{-1}$Mpc: This is the correlation length of faint ($B>22$) galaxies
(e.g., Roche \etal1993, Efstathiou 1995, Hudon \& Lilly 1996).
\item $r_0=5\point5\,h^{-1}$Mpc: This is the present-day QSO--galaxy correlation length
(Paper~I). Also the galaxy correlation length, derived
from bright, low-redshift galaxies has been measured by a variety of
authors (e.g., Seldner \& Peebles 1978, Davis \& Peebles 1983) to be in the
range $3\point4$ to $5\point5\,h^{-1}$Mpc (Koo \& Szalay~1984). 
\item $r_0=10\,h^{-1}$Mpc: This is an approximation to the clustering of galaxies 
around radio-loud QSOs. Yee \& Green (1987)
found correlation lengths of $8\point7$ and $17\point5\,h^{-1}$Mpc for radio-loud 
QSOs in the ranges $0\point3 < z < 0\point5$ and $0\point55 < z < 0\point65$
respectively. Using an enlarged data set, Ellingson \etal(1991)
again found $r_0\ge 15\,h^{-1}$Mpc for
$0\point3 < z < 0\point6$, so 10$\,h^{-1}$Mpc may be viewed as a lower limit. 
\endlist

For comparison, Loan, Wall \& Lahav (1997) measured $r_0\approx
18\,h^{-1}$Mpc for a mixed sample of radio galaxies and radio-loud
QSOs. Results from the FIRST radio survey are giving the lower value of
8--12$\,h^{-1}$Mpc (Cress \& Kamionkowsky 1998, Magliocchetti \etal1998), possibly
because the fainter flux limit is including many more starburst objects
than the Parkes and Greenbank surveys used in the Loan et al. paper.
Dalton \etal(1994) measured $r_0\approx 14\,h^{-1}$Mpc for the
galaxy cluster--cluster correlation length, but if we imagine that all
QSOs are found in galaxy clusters, a more appropriate comparison might
be with the cluster--galaxy cross-correlation, measured by Lilje \&
Efstathiou (1988) to be $r_0=8\point8\,h^{-1}$Mpc with
$\gamma=2\point2$.

From previous studies, a general picture has emerged of $\Bgg$ being
broadly similar to $\Bqg$ for radio-quiet QSOs.  Ellingson \etal(1991)
measured a QSO--galaxy correlation length of $5\point6\pm
3\point2\,h^{-1}$Mpc for their radio-quiet sample with no detectable
variation over the range $0\point 3 \le z \le 0\point 6$.  Boyle \&
Couch~(1993) detected no significant clustering around their optically
selected QSOs ($0\point 9 \le z \le 1\point 5$).  Though their result
($r_0 = 0\point 9 \pm 4\point0\,h^{-1}$Mpc) is consistent with a range
of amplitudes up to and including the present-epoch galaxy covariance,
it is inconsistent with the very rich environments measured by
Tyson~(1986) for radio-loud QSOs in the same redshift range.
Croom \& Shanks (1998) actually found a weak anti-correlation between $b_J<23$
galaxies and optically selected QSOs. They show that this can be
explained as an artifact caused by gravitational lensing, a conclusion consistent
with the earlier study of Ben\'\i tez \& Mart\'\i nez-Gonz\'alez (1997)
which used an even brighter magnitude limit of $b_J<20\point5$.

\begintable{2}
\caption {{\bf Table 2.}
  Covariance amplitudes and the corresponding correlation lengths 
  for $0\point3<z<0\point7$ QSOs and $V\le23$ galaxies under various assumed
  models of the clustering evolution, calculated from counts to one arcminute
  (equation~19).}
\centreline{
\vbox{
\tabskip=1em plus 2em minus .5em
\halign{\hfil#\hfil&\hfil#\hfil&\hfil#\hfil\cr
\noalign{\hrule}
\noalign{\smallskip} 
\noalign{\hrule}
\noalign{\medskip} 
Covariance Amplitude&Correlation length&Inconsistency\cr
&&with 1$\,$arcmin counts\cr
$A$ (at $1\,$arcmin)&$r_0$ ($h^{-1}$Mpc)&$\sigma$\cr
\noalign{\smallskip}
\noalign{\hrule}
\noalign{\medskip}
Observed\hfill&Derived\hfill&\cr
$0\point054\pm0\point020$&$2\point6\pm0\point6\ (\beta=1)$&--\cr
$0\point054\pm0\point020$&$3\point3\pm0\point6\ (\beta=0)$&--\cr
\noalign{\medskip}
Derived\hfill&Assumed\hfill&\cr
0$\point$606 ($\beta=1$)&10           &16\cr
0$\point$398 ($\beta=0$)&10           &10\cr
0$\point$207 ($\beta=1$)&5$\point$5  &5$\point$0\cr
0$\point$135 ($\beta=0$)&5$\point$5  &2$\point$7\cr
0$\point$033 ($\beta=1$)&2$\point$0  &0$\point$7\cr
0$\point$022 ($\beta=0$)&2$\point$0  &1$\point$1\cr
0$\point$000            &0           &1$\point$8\cr
\noalign{\medskip}
\noalign{\hrule}
} }
}
\endtable

In contrast, the results presented here are not consistent with
the canonical $r_0=5\point5\,h^{-1}$Mpc 
galaxy correlations. Also, they are greatly inconsistent with 
correlations as strong as have been claimed for radio-loud QSOs or the 
galaxy--cluster cross-correlation. We know, however that faint galaxies are 
not found in the same average 
environments as bright galaxies either. Combining the results from five 
separate determinations of $\Agg$, Efstathiou (1995) found a constant comoving 
correlation length $r_0=2\,h^{-1}$Mpc to be a good fit for faint 
($22 < b_J < 25\point5$) galaxies.  
A direct measurement of spatial clustering at moderate redshift
from the {\sl Canada-France Redshift Survey\/} (CFRS; Le F\`evre \etal1996) also 
gave the same value of $r_0=2\point0\,h^{-1}$Mpc, assuming no evolution in
comoving space between now and the mean redshift of the observations 
($z=0\point53$). These clustering amplitudes are consistent with the 
observations presented herein. As discussed by Le F\`evre \etal(1996),
the inconsistency between observations at low and intermediate redshifts
can be resolved either by assuming that a
different population of galaxies is being probed by faint surveys, or 
conversely if the one population has undergone very significant evolution
over the range $0<z<1$. The Le F\`evre \etal measurement of 
$r(z=0\point53)=1\point33\pm0\point09\,h^{-1}$Mpc is consistent with 
$r_0=5\,h^{-1}$Mpc if the evolution parameter, $\beta=-1\point6$.
This is a very rapid evolution.
A value of $\beta=-0\point667$ is predicted from $\Omega_0=1$ linear 
theory (Peebles 1988).

These results imply that radio-quiet QSOs apparently 
reside in environments consistent with those of the `normal' galaxy
population local to the QSO. 
We note however, that the observations are only 2$\sigma$
inconsistent with there being no correlation at all, i.e., $r_0=0$, 
and our detection of clustering is not as secure as that in
large galaxy redshift surveys such as the CFRS. 
Any 2D clustering results are equally consistent with an arbitrarily large 
correlation if we are prepared to accept the associated rapid evolution. 

The covariance amplitudes for $0\point3<z<0\point7$ QSOs and
$V\le23$ galaxies are summarised in Table~2. (The actual observed
counts may be found in Table~3.) In the first two rows,
the observed amplitude is determined from counts at less than
1\,arcmin and converted to the correlation length,
$r_0$.  For the rest of table, assumed spatial-clustering
models are converted to $A_{qg}$ via equation 14 for comparison.  The
final column gives the Poisson deviation required on the 1\,arcmin
galaxy counts to produce the suggested covariance amplitude. The most
probable correlation lengths for non or slowly evolving clustering are
around $3\,h^{-1}$Mpc. Correlation lengths as great as
$r_0=5\,h^{-1}$Mpc become acceptable only if there is very rapid
evolution in the clustering amplitude.  Clustering comparable to that
reported for radio-loud QSOs is strongly ruled out.

\subsection{Subdivided data-sets}
In order to investigate the parameters in the evolution model and
to reduce the effect of clustering dilution along the line of sight,
we sub-divided the sample in redshift. We also repeated the
analysis for two sub-samples defined by absolute magnitude.

Observed 1\,arcmin pair counts for the low ($0\point3<z<0\point5$) and high-redshift
($0\point5<z<0\point7$) samples are given in Table~3 and the cross-correlation
functions are plotted in Fig.~5.  Results from these samples
demonstrate that most of the clustering seen in the full sample comes from
the low-redshift fields, but that both sub-samples are consistent with
either comoving or stable clustering with an $r_0$ of
2--3$\,h^{-1}$Mpc.  The $z<0\point5$ data are best fit by $r_0 =
3\point6\,h^{-1}$Mpc, for virialized, stable clustering and $r_0 =
2\point9\,h^{-1}$Mpc for comoving clustering, assuming a
$\gamma=1\point8$ and the $A_{qg}=0\point0028$ least-squares
determination of the amplitude. The slightly weaker
$A_{qg}=0\point0026$ calculated from the one arcminute counts would
give correlation lengths shorter than those quoted above by
$0\point1\,h^{-1}$Mpc.  The galaxy magnitude limits again
correspond to $M^*-5\log h+1$ at the median redshift of the
sub-sample. Inspection of Fig.~3 shows that in both cases the $N(z)$
distributions peak within an appropriate redshift range.

\beginfigure{5}
  \vfill
  \centreline{\psfig{figure=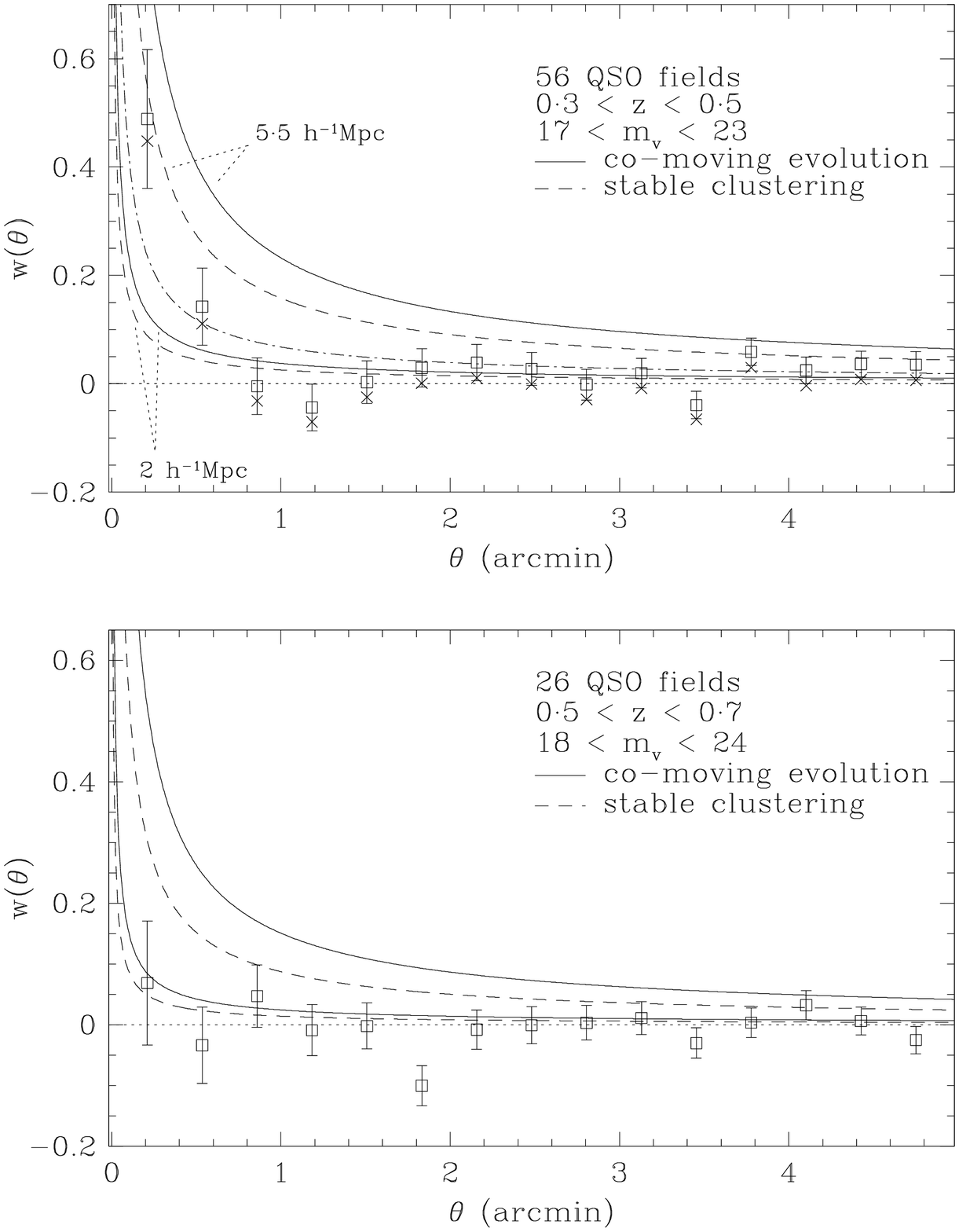,width=\hsize}}
  \caption{{\bf Figure 5.}
     Cross correlations for $z<0\point5$ and $z>0\point5$ data plotted 
     separately. In the upper, low-redshift panel, crosses are the raw, observed correlation and
     open squares are corrected for the integral constraint ($11\point0 A_{qg}$). 
     No correlation is detected in the higher-redshift sample, so no
     integral constraint is applicable. The dot-dashed curve in the upper
     panel is the measured amplitude of $A_{qg}=0\point0026\pm0\point001$, which 
     gives $r_0=2\point8\,h^{-1}$Mpc ($\beta=1$). A least-squares best fit 
     to the observations gave the consistent value, $A_{qg}=0\point0028$.   
     Comparison curves are for $5\point5$ and $2\,h^{-1}$Mpc as in Fig.~4.}
  \vfill
\endfigure

\begintable{3}
\caption {{\bf Table 3.}
  Galaxy and random point raw counts at less than one arcminute
  from the QSO. $\sigma$ is the Poisson
  significance of $QG$'s difference from $QR$. Note that in two cases this is
  is a very small deficit rather than an excess, formally signifying an anti-correlation.
  After inclusion of the correction for `detection success ratio' described in Section~2.5,
  these become a small, but insignificant positive correlation.
}
\centreline{
\vbox{
\tabskip=1em plus 2em minus .5em
\halign{\hfil#\hfil&\hfil#\hfil&\hfil#&#&#\cr
\noalign{\hrule}
\noalign{\smallskip} 
\noalign{\hrule}
\noalign{\medskip} 
QSO sub-sample    & Galaxy magnitudes&$QG$&  $QR$& $\sigma$\cr
\noalign{\smallskip}
\noalign{\hrule}
\noalign{\medskip}
$0\point3<z<0\point7$&$17<V<23$&1008&948&1$\point$9\cr
$0\point3<z<0\point5$&$17<V<23$&691&640&1$\point$9\cr
$0\point5<z<0\point7$&$17<V<24$&725&729&0$\point$15\cr
$M_V>-22$&$17<V<23$&607&548&$2\point4$\cr
$M_V<-22$&$17<V<23$&401&403&0$\point08$\cr
\noalign{\medskip}
\noalign{\hrule}
} }
}
\endtable

\beginfigure{6}
  \vfill
  \centreline{\psfig{figure=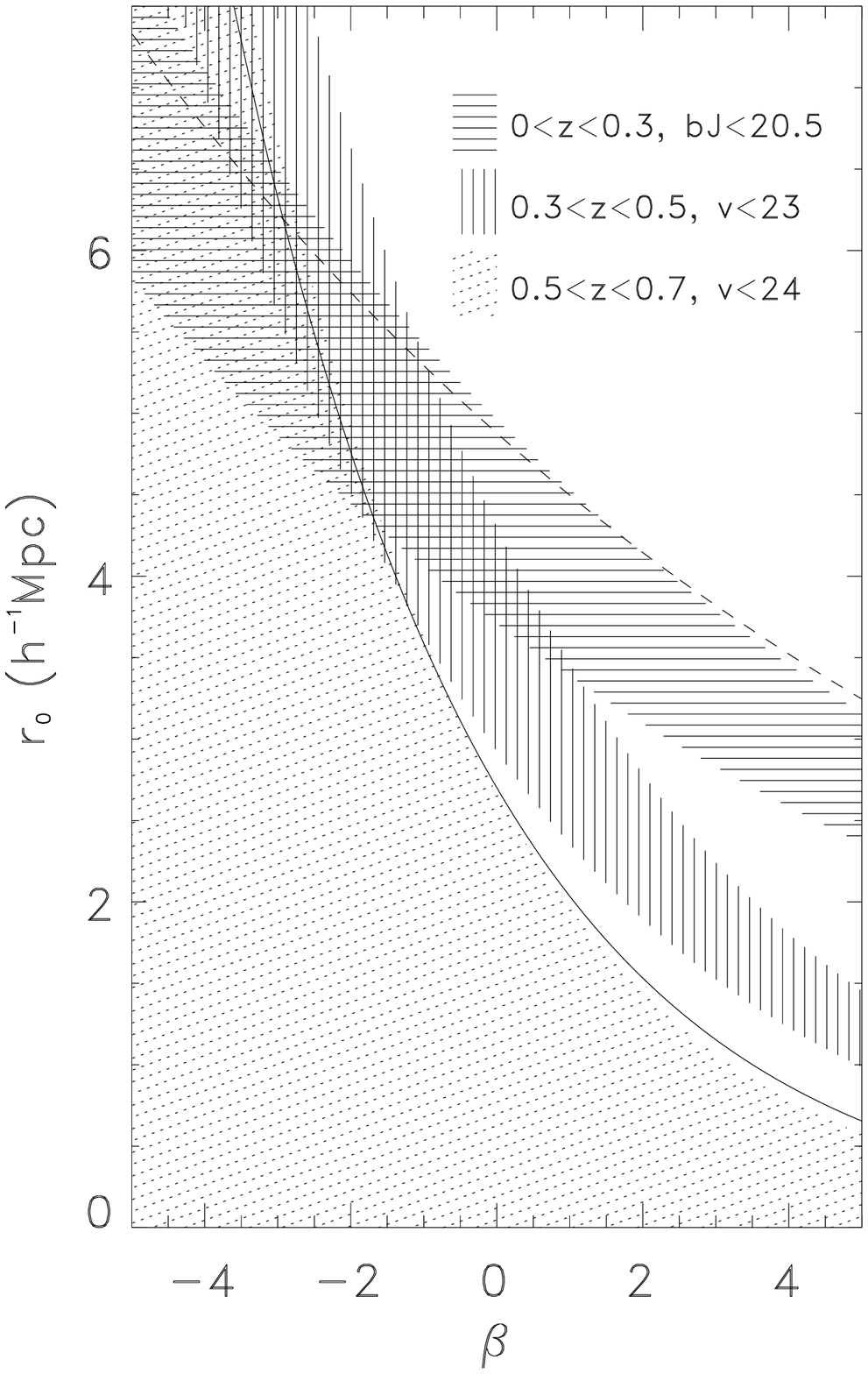,width=\hsize}}
  \caption{{\bf Figure 6.}
    Values of $r_0, \beta$ which would be consistent with the present
    and previous observations. Shaded regions show $1\sigma$ limits
    on our observations of $\Aqg$. The solid line represents $\xi_{gg}$
    from the CFRS sample and the dashed line $w(\theta)_{gg}$ from
    the APM Galaxy Survey.}
  \vfill
\endfigure

\beginfigure{7}
  \vfill
  \centreline{\psfig{figure=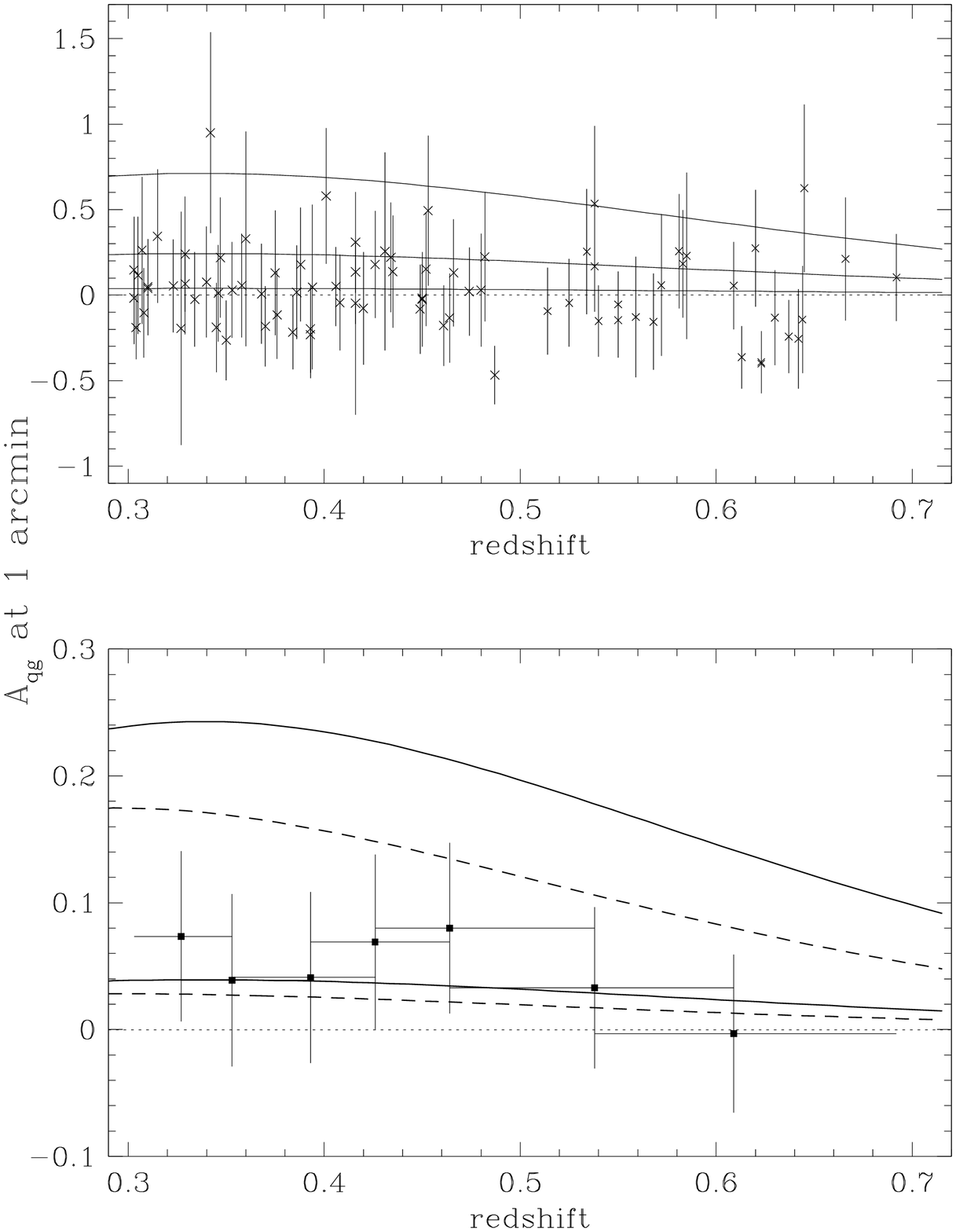,width=\hsize}}
  \caption{{\bf Figure 7.}
    Observed QSO--galaxy cross-correlation amplitude at one arcminute
    using $V\le23$ galaxies. The two panels show the same
    data with an axis re-scaled for clarity. In the top panel, curves
    representing 10, $5\point5$ and 2$\,h^{-1}$Mpc comoving correlation
    functions are shown. In the lower panel, only $5\point5$ and 2$\,h^{-1}$Mpc
    are visible. Solid curves are for constant
    comoving evolution and dashed for constant proper correlation length.
    Filled boxes are each the average of eleven fields and the error bars
    give the one sigma standard deviation in each subsample.}
  \vfill
\endfigure

We can also combine the correlation length derived for the galaxy
environments of $z<0\point3$ QSOs in Paper~I with those derived here
to give us a large baseline over which to study the evolution of the
clustering.  A single measurement of the correlation defines 
a curve in the $r_0, \beta$ plane giving all possible value pairs consistent 
with the observation. Fig.~6 shows the regions of this plane 
bounded by the $1\sigma$ uncertainties on our clustering amplitude
determinations, $0<z<0\point3$, $0\point3<z<0\point5$ and $0\point5<z<0\point7$.
The high-redshift sample only defines an upper limit. 
Also plotted are two curves taken as examples of $\xi_{gg}$
measurements. The dashed line is for $b_J<20$ galaxies from the APM
Galaxy Survey (Baugh 1996) which have a median redshift, $z_m=0\point13$.
The solid line is for the CFRS ($I<22\point5$), which has median
redshift, $z_m=0\point53$. These are not necessarily absolutely comparable to 
our amplitudes, since although they cover very similar redshift ranges, they have 
different magnitude limits and used slightly flatter power-laws to fit
the clustering amplitude. They are however in reasonably good agreement.

The various observations may thus be reconciled if there has been 
moderately strong evolution in the clustering amplitude, $\beta < -1$
for both $\xi_{qg}$ and $\xi_{gg}$. Such a conclusion must assume that
the same single population is being sampled in each experiment.

Alternatively, rather than summing the data over a predefined redshift
range, in Fig.~7 the observed amplitude at one arcminute is
plotted as a function of redshift for each QSO individually.  The top
panel shows the amplitude for each of the 82 fields with their
individual error bars and three constant comoving correlation length
models. In the lower panel, the $A_{qg}$ axis is re-scaled for
clarity. Filled boxes are each the average of eleven fields,
thereby adaptively sampling the redshift range according
to the availability of data.  The $r_0=5\point5\,h^{-1}$Mpc and
$2\,h^{-1}$Mpc, comoving and stable clustering models are also plotted
in the lower panel. 
This demonstrates that the summed data in the redshift
slices are representative and are not strongly dependent on any
single, or few, fields. There are significantly more positive points
than negative in the $z<0\point5$, $V\leq23$ data.
When a similar plot is produced with a limiting galaxy 
magnitude $V\le24$, which we consider more appropriate for 
$0\point5< z < 0\point7$ QSOs, the data points are almost exactly evenly split
around the $\Aqg=0$ axis at all redshifts, giving agreement with the zero 
correlation seen in the lower panel of Fig.~5. 

The plot also shows that the predicted change in amplitude over the
redshift range studied is small and the curves for different values of
the parameter $\beta$ are similar in shape. An order of magnitude
increase in the number of QSO fields observed would be required to
reliably distinguish between these two clustering evolution
schemes. 

Finally, the clustering amplitude as a function of QSO absolute magnitude
is shown in Fig.~8. Amplitudes are here plotted in terms of the 
spatial covariance to compensate for the different projection
effects on QSOs of the same absolute magnitude, but different redshifts.
The amplitude appears to decrease with increasing QSO luminosity. 
This is markedly different to the rapid increase in cluster richness 
with QSO luminosity seen by Yee \& Green (1987)
for radio-loud QSOs. Caution must be exercised when assigning any 
significance to the apparent variation seen in Fig.~8. Figs~7 and 8 are not independent
since luminosity and redshift are related via the QSO luminosity
evolution. Indeed, from Fig.~1 it is clear that
80 per cent of the $M_V-5\log h>-22$ QSOs are also in the $z<0\point5$ 
sub-sample. 
What appears to be a correlation with luminosity may simply be the selection 
function being influenced by the evolving QSO luminosity
function. Allowing for this, there is no evidence in the current data for
a dependence on luminosity which is independent from that with redshift.

\beginfigure{8}
  \vfill
  \centreline{\psfig{figure=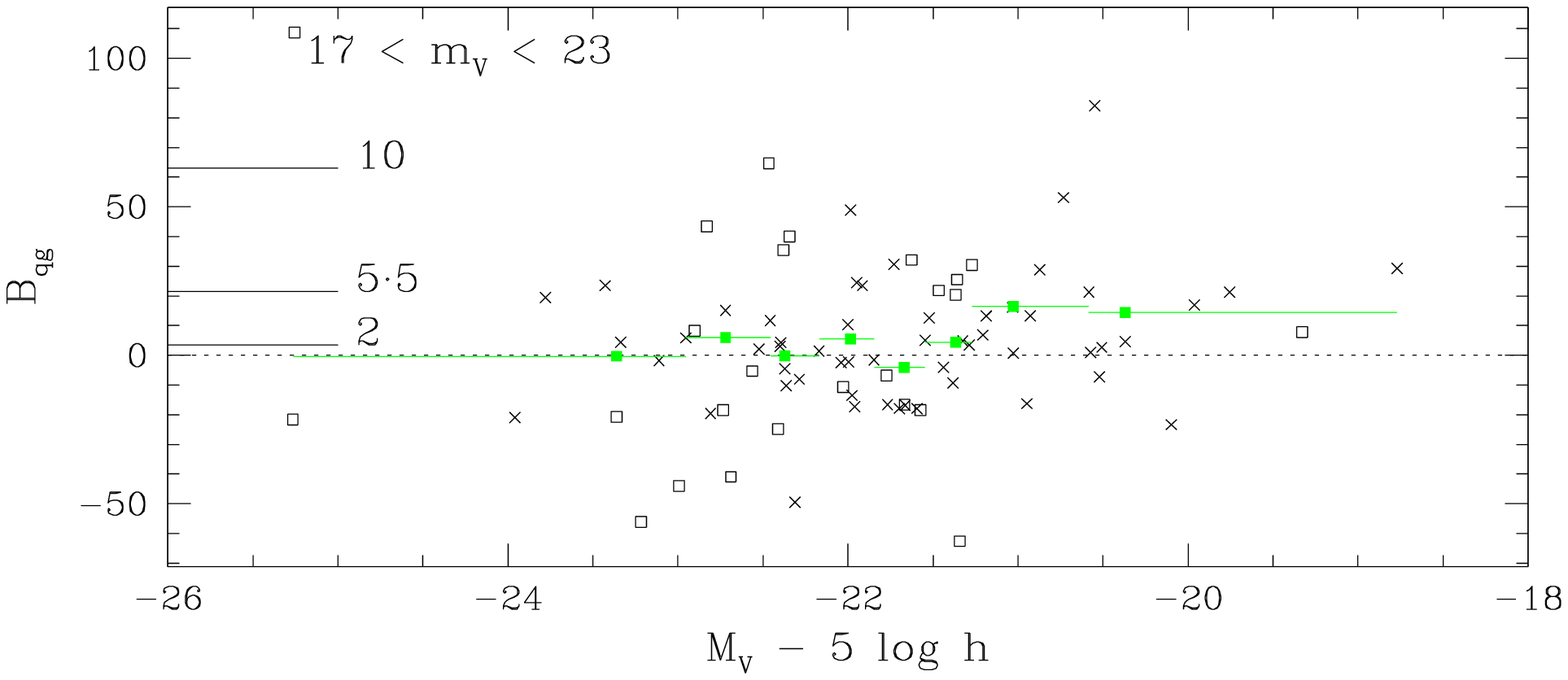,width=\hsize}}
  \caption{{\bf Figure 8.}
    QSO--galaxy spatial covariance amplitude, assuming $\beta=1$, as a 
    function of
    QSO absolute magnitude. Crosses are $z<0\point5$, open boxes $z>0\point5$.
    Horizontal bars on the left of the plot indicate the amplitudes
    equivalent to correlation lengths, $r_0=2$, $5\point5$ and 10$\,h^{-1}$Mpc.
    $M_V>-22$ is almost a subset of $z<0\point5$, so trends in $\Bqg$ from 
    this plot and Fig.~7 are not independent.}
  \vfill
\endfigure

\subsection{Comparison with QSO and galaxy autocorrelations}
In the previous section we showed that, at intermediate redshifts,
$0\point3<z<0\point7$, the inferred clustering scale length for faint galaxies 
around QSOs, $r_0=2\,h^{-1}$Mpc was similar to that derived from the
auto-correlation function of faint galaxies, and thus, in the simplest
interpretation, the QSOs and faint galaxies might exhibit the same bias.
In contrast, the most recent measurements of the QSO correlation function 
are consistent with a constant comoving amplitude  $r_0=5\,h^{-1}$Mpc 
at $z\le2\point2$ (Croom \& Shanks 1996), suggesting QSOs are strongly
biased with respect to the galaxy population if the galaxy scale
length is $r_0=2\,h^{-1}$Mpc.
However, the measurement of the QSO correlation length is dominated
by QSOs with $1<z<2$, with relatively few QSOs at $0\point3<z<0\point7$ and certainly
too few to rule out $r_0=2\,h^{-1}$Mpc for the QSO correlation length
at these redshifts.  

If the galaxy correlation length were to increase again at $z>1$ then the
QSO and galaxy distribution could be reconciled with the same value for
the bias parameter (albeit redshift dependent).  
There is some evidence that the galaxy correlation must increase again
at higher redshift;
angular correlations for HDF galaxies with photometric redshifts shown a rapid
increase at $z \sim 2$ (Magliocchetti \& Maddox 1999, Arnouts et al 1999), and
Giavalisco \etal (1998) measure $r_0=6\,h^{-1}$Mpc for galaxies at $z\sim3$.
Alternatively, the strong clustering seen for both QSOs and 
the Giavalisco \etal galaxies may simply be highlighting the danger
of generalising from any one type of galaxy to another when we do not
understand galaxy formation and how the classes are inter-related.
Such discrepancies are seen also for `normal' galaxies, even within a single,
optically selected survey. For example, both the luminosity functions (Lilly \etal1995)
and correlation functions (Le F\`evre \etal1996) of red and blue galaxies
in the CFRS show significantly different evolution over the redshift 
range $0<z<0\point8$ and in comparison to the present epoch determinations.

\subsection{Future work}
It is clear that we do not yet understand the cluster environments of
QSOs with the same precision as we do for galaxies. New wide-field CCD
cameras on large-aperture telescopes will be able to make some
contribution to these studies, but the broad range in redshift of
galaxies in a simply flux-limited sample places a significant limit on
the sensitivity of this approach.  The ever improving availability of
redshift data for faint galaxies will go some way to allowing detailed
studies of QSO environments, though even the upcoming generation of
sky surveys will not probe to sufficiently faint limits to obtain
redshifts for typical $M^*$ galaxies at the redshifts investigated
here. Photometric redshifts obtained from wide-field imaging in
several pass-bands may prove the most efficient method to
further the work described here. The potential for photometric redshifts
at $z<2$ has been proved through recent interest in the technique
spurred by the Hubble Deep Field (e.g., Fern\'andez-Soto, Lanzetta \& Yahil 1998).

\section{Summary}
Clustering of $V<23$ galaxies around X-ray-selected, $0\point3<z<0\point7$  QSOs has been detected 
with a ${>}2\sigma$ significance. The amplitude of the correlation function is 
small, $A_{qg}(1^\circ)=0\point0020\pm0\point0008$, implying a spatial clustering of amplitude
$r_0\approx3\,h^{-1}$Mpc. 
Though not a secure detection of clustering around X-ray selected QSOs,
it is a highly significant rejection of rich cluster environments similar
to those found for radio loud QSOs (Yee \& Green 1987), and of the
cluster environment around present day normal galaxies.

When the sample was subdivided at $z=0\point5$, the correlation was found to be stronger in the
$z<0\point5$ data. No significant excess of QSO--galaxy pairs over the 
mean background counts was observed at $z>0\point5$.  The best-fitting amplitude for the low-redshift 
sample was $A_{qg}(1^\circ)=0\point0026\pm0\point001$, which corresponds to $r_0\approx3\,h^{-1}$Mpc
if we assume there has been little evolution in the clustering since $z=0\point5$.

These measured correlation amplitudes are in good agreement with
measurements of the faint-galaxy correlation function (e.g., Roche \etal1993, Le F\`evre \etal1996).
Taken in conjunction with the results of Paper~I, this implies that
QSOs are likely to be populating average galaxy environments at all redshifts $z<0\point5$.
Though no clustering was detected for the $z>0\point5$ subsample or in the
high-redshift sample of Boyle \& Couch (1993), these null results too are consistent 
with the same conclusion.  This is in very marked contrast to the rapid evolution
in the richness of galaxy clusters associated with radio-loud QSOs,
which by $z\sim0\point6$ are typically found in Abell richness class 1 clusters
(Yee \& Green 1984, Ellingson \etal1991).  While further work remains to be done, the
results here indicate that radio-quiet QSOs may not be strongly biased
with respect to the galaxy population.  This has importaint, and
potentially positive, consequences for studies of large-scale structure from the 
forthcoming large QSO redshift surveys.  

\section*{Acknowledgements}
RJS acknowledges the financial support of the United Kingdom Particle
Physics and Astronomy Research Council and would also like to thank
the Anglo-Australian Observatory
for their generous hospitality during a one-year visit to 
their Epping laboratory, where much of this work was conducted.

\bye